\documentclass[galley,grl]{agu2001}
\usepackage{graphicx}
\authorrunninghead{S. BHATTACHARYYA AND R. NARASIMHA}
\titlerunninghead{MONSOON RAINFALL AND SOLAR ACTIVITY}
\authoraddr{S. Bhattacharyya, R. Narasimha, Engineering Mechanics Unit, 
Jawaharlal Nehru Centre For Advanced Scientific Research,
Jakkur, Bangalore 560064, India,
roddam@caos.iisc.ernet.in}
\begin{document}
\title{ Indian monsoon rainfall is higher in epochs of higher solar activity: a wavelet cross-spectral analysis}
\author{S. Bhattacharyya and R. Narasimha}
\affil{Engineering Mechanics Unit, 
Jawaharlal Nehru Centre For Advanced Scientific Research,
Jakkur, Bangalore-560064, India}
\begin{abstract}
Using the Morlet continuous 
wavelet transform on data over the period 1871-1990, it is found that the global wavelet cross spectra between two 
solar activity indices and seven major Indian monsoon rainfall time series show significant
power around the period of the 11 year solar cycle, passing the $\chi^2$ test of significance proposed by Torrence and Compo (1998) at levels exceeding $95\%$ in 10 out of the 14 cases studied. Furthermore two distinct epochs are found in the computed cross-spectrum, the transition between them occurring around the decade 1915-1925, beyond which solar activity parameters show a significant increase. By comparison between selected periods of three cycles in solar activity
in each of the two epochs, it is shown that the average rainfall is higher in all seven rainfall indices during periods of greater solar activity, at {\it z}-test confidence levels of $95\%$ or greater in three of them.
\end{abstract}
\begin{article}
\section{Introduction}
There are numerous studies of the possible influence of solar
activity on terrestrial climate. Although solar-terrestrial connections have been somewhat sceptically received by  much of the meteorological community, the growing availability of paleo-climate indicators using novel measurement techniques, and the discovery of the prominent solar activity period of 11 years in varied climate records
using new mathematical tools, are producing increasingly strong evidence for possible association between solar processes and terrestrial weather and climate indices (Beer {\it et al.}(1990), Friis-Christensen and Lassen (1991), Lassen and Friis-Christensen (1995), Haigh (1996, 1999, 2001), Labitzke and Loon (1997), Mehta and Lau (1997)). Doubts about effects on shorter time scales (of the order of 100y) seem however to remain. Earlier studies of possible connections between Indian monsoon rainfall and solar activity (Jagannathan and Bhalme (1973) and Jagannathan and Parthasarathy (1973)), using correlation and power spectral analysis of the rainfall distribution in $48$ meteorological stations spread all over India, reported presence of the 11-year sunspot cycle at significance levels of $95\%$ or higher in $5$ of them.

The possibility of shedding light on the variability of the Indian monsoon using wavelet techniques (Kailas and Narasimha (2000), Narasimha and Kailas (2001)), and the idea that the tropics can amplify a small radiant flux signal to a relatively large and dynamic climate change elsewhere in the world as well (Haigh (2001), Visser {\it et al.} (2003)), serve to provide further motivation for the present work. In a recent study by Neff {\it et al.} (2001), strong coherence 
between the solar variability and the monsoon in Oman in the period between 9 and 6 kyr B.P. has been reported.

There has been a recent surge in interest in incorporating wavelet techniques for climate signal detection and analysis, as can be seen in the works of Lau and Weng (1995) and Torrence and Compo (1998). In particular, Narasimha and Kailas (2001), analysing Indian monsoon rainfall, identified six 'quasi cycles', with average periods of 2.92, 5.8, 11.4, 19.9, 34.2 and 79.8 y. These numbers suggested the presence of several super- and sub-harmonics of the basic sun-spot period of about 11 y. Iyengar and Raghukantha (2004) have identified intrinsic modes with periods virtually identical to those observed by Narasimha and Kailas (2001). 

In continuation of this on-going research, we now present an analysis of the association between indicators of solar activity and Indian monsoon rainfall, using the continuous wavelet transform method with the Morlet wavelet function. 
\section{The data analysed}
Seven annual area-weighted rainfall time series for the period 1871-1990 have been considered for the analysis, namely, all India summer monsoon rainfall (AISMR), north east India (NEI), north west India (NWI), central north east India (CNEI), west central India (WCI), peninsular India (PENSI) and the homogeneous Indian monsoon (HIM)(Parthasarathy{\it et al} (1995)). 
The choice of these regions is based on several well defined criteria including primarily a considerable degree of spatial coherence. The HIM region covers the central and north-western parts of India amounting to $55\%$ of the total land area of the country, and may be seen as the most characteristic index of the component of Indian rainfall dominated by the dynamics of the south west monsoon. As it is well known that the NEI region shows in many ways an anti-correlation with the HIM region (Parthasarathy {\it et al.} 1993), we present more detailed results specifically for these two regions.

The range of scales over which these rainfall data can provide useful information on temporal variability is limited at one end by resolution, since not more than 12 points per year are available, and at the other end by the limited length of data stretch of 120 years. For the present study we have found annual rainfall to be the most appropriate rainfall index to use. The solar indices under study are sunspot number index and group sunspot number. Sunspot areas have also been studied, but have been found to provide no new information. The sunspot index data have been obtained from Rai Choudhuri (1999) and Fligge {\it et al.} (1999), and the data for group sunspot number from the NOAA ftp site {\it ftp://ftp.ngdc.noaa.gov/STP/SOLAR\_DATA}. 

We must note one important difference between the rainfall and
sunspot data. The former are cumulative, the monthly
data being a sum of the daily data. That is not the case for
sunspot data, since individual sunspots live typically for
several days. All monthly data of sunspots are usually compiled
by taking averages of daily data over a month. The monthly
average sunspot number plotted against time does not appear very
smooth (see, for example, the website http://science.msfc.nasa.gov/ssl/pad/solar/image/zurich.gif).
As our aim here is to study possible correlations of the
rainfall data with solar processes with time scales of order years to decades, we
filter out fluctuations in the sunspot data at small time scales
by using yearly sunspot and group sunspot numbers.
\section{Wavelet Cross Power Spectrum }
We use the Morlet wavelet function
\begin{equation}
\psi(\eta)=\pi^{-1/4}e^{i\omega_{0}\eta}e^{-\eta^{2}/2},
\end{equation}
where $\omega_{0}$ is a nondimensional frequency, taken equal to 6
in order to satisfy the wavelet admissibility condition, and
$\eta$ is a nondimensional time parameter.
For a discrete sequence $R_n$, $n=0,..,(N-1)$, the continuous wavelet transform $W_{n}^{R}(s)$ 
is defined as a convolution of $R_n$ with a scaled and translated 
version of the wavelet function $\psi(\eta)$, as given by the expression
\begin{equation}
W_{n}^{R}(s)=\sum_{n^{'}=0}^{N-1}R_{n^{'}}\psi^{\star} [ \frac{(n^{'}-n)\delta t}{s} ],
\end{equation}
where $^{\star}$ denotes the complex conjugate, and $\delta t$ is the (sampling) time interval between two consecutive points in the time series. The wavelet function at each scale $s$ is normalised to have unit energy, so that 
\begin{equation}
\hat\psi(s\omega_{k})=\left ( \frac{2\pi s}{\delta t}\right )^{1/2}\hat\psi(s\omega_{k}),
\end{equation}
\begin{equation}
\int_{-\infty}^{\infty}\mid \hat\psi(\omega^{'})\mid^{2}d\omega^{'}=1,
\end{equation}
where $\hat \psi$ is the Fourier transform of $\psi$.
The wavelet power spectrum of $R_n$ is given by the convolution 
$W_n^{R}(s)*\left[W^{R(s)}_{n}\right]^{\star}$ and the wavelet power is given by the magnitude
$\mid W_n^{R}(s)*\left[W_{n}^{R}(s)\right]^{\star} \mid$.
For the wavelet cross power spectral analysis we utilize the 
easy-to-use toolkit, including statistical significance testing, 
as outlined by Torrence and Compo (1998).
The cross wavelet spectrum $W_{n}^{RS}(s)$ between two time
series $R_n(t)$ and $S_n(t)$, with the respective wavelet transforms 
$W_{n}^{R}(s)$ and $W_{n}^{S}(s)$, may be defined as
\begin{equation}
W_{n}^{RS}(s)=W_{n}^{R}(s)\left[ W_{n}^{S}(s)\right ]^{\star}.
\end{equation}
The cross wavelet power is 
$\mid W_{n}^{RS}(s) \mid$. 
Torrence and Compo (1998) derive the confidence levels for the cross wavelet power
from the square root of the product of two $\chi^2$ distributions.
In the test, a peak in the wavelet power spectrum is considered to be a 
true feature, with a certain percentage confidence, if the peak is significantly above a background or reference spectrum
given by
\begin{equation}
P_k=\frac{1-\alpha}{1+\alpha^2 -2\alpha\cos(2\pi k/N)}.
\end{equation}
Here $k$ is the frequency index, and $\alpha=(\alpha_1 +\sqrt \alpha_2)/2$ where $\alpha_1$ and $\alpha_2$ are the lag-1 and lag-2 autocorrelation coefficients of the process under consideration. For a white noise background spectrum $\alpha=0$. 
If the two time series have background spectra
given respectively by $P_{k}^{R}$ and $P_{k}^{S}$, then the cross wavelet power 
distribution will be given by
\begin{equation}
\frac{ \mid W_{n}^{RS}(s) \mid}{\sigma_{R} \sigma_{S}} ==> 
\frac{Z_{\nu}(p)}{\nu} \sqrt{P_{k}^{R} P_{k}^{S} },
\end{equation}
where $\sigma_{R}$ and $\sigma_{S}$ are the standard 
deviations of $R$ and $S$ respectively, $\nu$ is the number of degrees of freedom with which
$\chi^2$ is distributed, $p$ denotes the level of confidence and
$Z_{\nu}(p)$ denotes the value of the $\chi^2$ distribution with $\nu$ 
degrees of freedom at the confidence level $p$. For the complex Morlet wavelet $\nu=2$ . 
\section{Results}
A plot of the annual time series of the four rainfall and two 
solar indices under consideration is shown in figure 1. 
The rainfall time series appear irregular and random, 
while the solar indicators have a clearly cyclic character.

The results of the present analysis are presented in the form of colour-coded contour maps (see figures 2 and 3) of 
wavelet cross power spectra as functions of time and Fourier period (henceforth refered to as period) respectively for HIM and NEI rainfall. Outlined on these graphs are thick contours enclosing regions where wavelet cross power exceeds $95\%$
confidence levels, with respect to the reference spectra mentioned above. The cones of influence within which edge effects become important are also indicated by dashed lines in the figures. 
 
Figures 4 and 5 show several global wavelet power spectra as functions of period for HIM and NEI rainfall.  
For HIM rainfall (figure 2), a wavelet cross power of noticeably high magnitude is observed
at the middle of the period range 8-16 years. An integration over time of this wavelet cross power, which gives the global wavelet cross power spectrum, is shown in figure 4. It can be seen clearly from this figure that
the cross wavelet power crosses the $95\%$ confidence line at 
a period of around 11 years. Circles and squares 
respectively denote the contributions of the individual global wavelet power
of the rainfall and the sunspot time series  to the global spectrum.
It is seen that the contribution of the 8-16 year period band is  
$16.8\%$ for the total rainfall  and $82.8\%$ for the sunspot index.

Figures 3 and 5 show plots similar respectively to figures 2 and 4 but for NEI 
rainfall and group sunspot number. Significant power is observed in the 8-16 year period band in this case also.
The global cross wavelet power spectrum also
crosses the $95\%$ confidence line in the period range 8-16 years (figure 5); the corresponding cross power (see figure 2) is however lower than that
for HIM. The highest cross power is observed during periods 1890-1905 and 1915-1965
in the former case, and during 1880-1920 and 1935-1980 in the latter case.
In both cases, therefore, there is strong indication of 
two relatively distinct epochs, the transition from one to the other 
occurring around 1915-1925. As may be seen from Figure 1, solar activity generally shows an increase beyond this decade, suggesting that higher solar activity is associated with the higher cross spectrum. 

Similar analyses have been carried out for the other 5 rainfall indices as well, and the results are summarised in Table 1.
It will be seen that, except for NEI and PENSI rainfall, all the other cases show up regions of cross power at confidence levels exceeding $95\%$ at the 8-16 year period band.

From figure 1, two epochs of low and high 
group sunspot number can be identified, this transition also occurring around 1915 to 1925. We select here one time interval in each of the two epochs, namely 1878-1913 and 1933-1964 respectively,
during each of which three complete solar cycles are present (considering the period from one minimum to the next as one complete cycle). Table 1 lists the annual rainfall means $\mu_1$ and $\mu_2$ respectively for the two periods 1878-1913 and 1933-1964; both the epochs and the corresponding means are shown in figure 1. The null hypothesis that the difference in mean annual rainfall between these two periods is zero is rejected at the maximum confidence levels listed in Table 1 using a one-tailed {\it z}-test (Crow {\it et al.} 1960). Thus the mean annual rainfall during 1933-64 (higher solar activity) is everywhere higher than that during the period (1878-1913) of lower solar activity. However the confidence levels are $95\%$ or higher in 3 cases out 7, including AISMR and HIM, and reaching $99\%$ in WCI. At the other extreme, it is a low $75\%$ in NEI and NWI. 

The ratio of wavelet power present in the period 8-16 years to the total rainfall wavelet power is also presented in table 1. So are the ratios of cross wavelet power between the rainfall and group sunspot number in the 8-16 year band to the total cross wavelet power. In the case of HIM and WCI, which show the solar-monsoon link at the highest confidence levels, the cross spectrum is also high (about $56\%$), and the contribution from the 8-16 y band is nearly $17\%$. It will be seen from Table 1 that, in general, significant effects on rainfall go with significant levels of cross power. The exception is the arid region NWI, where the lower significance levels for the differences in rainfall are due to the large relative standard deviation.
\begin{table}[h]
\vskip 1pc
\label{tab1}
\caption{\label{tab1}  Confidence levels  and $\%$ wavelet power }
\begin{tabular}{l  l  l  l  l  l  l  l} 
\hline
Region & $\mu_1$ & $\mu_2$ & \% con- & $\%$ power & $\%$ cross & $\%$ conf.,\\
 
         & mm \tablenotemark{a} & mm \tablenotemark{b} & fidence \tablenotemark{c} &  rainfall \tablenotemark{d} & power \tablenotemark{e} & cr-power \tablenotemark{f}\\
\hline
AISMR & 853.5 & 883.0 & 95   & 10.2 & 49.9 & 98\\ 
HIM   & 858.6  & 916.5 & 96.8& 16.8 & 56.7 & 99\\
WCI   & 1067.2 & 1145.7 & 99 & 16.9 & 56.4 & 98\\
PENSI & 1140.5  & 1183.2 & 85& 10.4 & 47.4 & 92.5\\
CNEI  & 1204.2 & 1235.4 & 80 & 7.7  & 45.7 & 95\\
NWI   & 542.0 & 565.1 & 75   & 15.3 & 55.9 & 99\\
NEI   & 2071.8 & 2100.2 & 75 & 8.6  & 45.8 & 98\\
\hline
\end{tabular}
\tablenotetext{a}{Mean rainfall over three cycles of low solar activity, 1878-1913.} 
\tablenotetext{b}{Mean rainfall over three cycles of high solar activity, 1933-1964.}
\tablenotetext{c}{Confidence level at which $\mu_1-\mu_2$ differs from zero by the {\it z-}test.}
\tablenotetext{d}{Contribution to total rainfall from 8-16 y band.}
\tablenotetext{e}{Contribution to cross-spectrum power from 8-16 y band.}
\tablenotetext{f}{Approximate maximum confidence level at which cross spectrum exceeds values for reference spectrum over a continuous period of at least 10 y.}

\end{table}
Incidentally the present results demonstrate the advantages of 
the wavelet approach, as compared with classical correlation/power 
spectral density methods: (i) wavelets permit identification of 
epochs during which correlations at different significance levels may 
have prevailed; (ii) wavelet methods allow us to take account of slight 
variations in the effective period or scale ('meandering') of the 
effect of a given forcing (as may be seen from the regions of high cross spectra in figures 2 and 3), such meandering being presumably the result of the nonlinear interactions between different modes of the system. 
\section{Conclusions}
The present study, involving two solar index time series and 
seven Indian rainfall time series and using wavelet cross spectral
density analysis as outlined by Torrence and Compo (1998),
reveals considerable power in the global cross power spectra 
around the 11 year solar cycle period for all the indices considered. 
In particular the global cross power spectra for AISMR, WCI and HIM rainfall with the group sunspot number reveal 
a significant peak at the 11 year period at confidence levels of $95\%$ or higher. Greater solar activity seems to be associated in all cases with greater rainfall, although at significance levels that are distinctly high in 3 and lower in 4 out of 7 cases studied. This regional variation is not inconsistent with the simulations of Haigh {\it et al.} (2004), which suggest that a major effect of higher solar activity may be a displacement in the Hadley cell. Such a displacement,
depending on its magnitude, can have different effects on rainfall in different regions.

\begin{acknowledgments}
The authors would like to thank Prof. A Rai Choudhuri, of the Physics Department of Indian Institute of Science, for his help on the solar data. The authors are grateful to the Centre for Atmospheric and Oceanic Sciences of the Indian Institute of Science for their continued hospitality.
\end{acknowledgments}


\begin{figure}
\noindent\includegraphics[width=20pc]{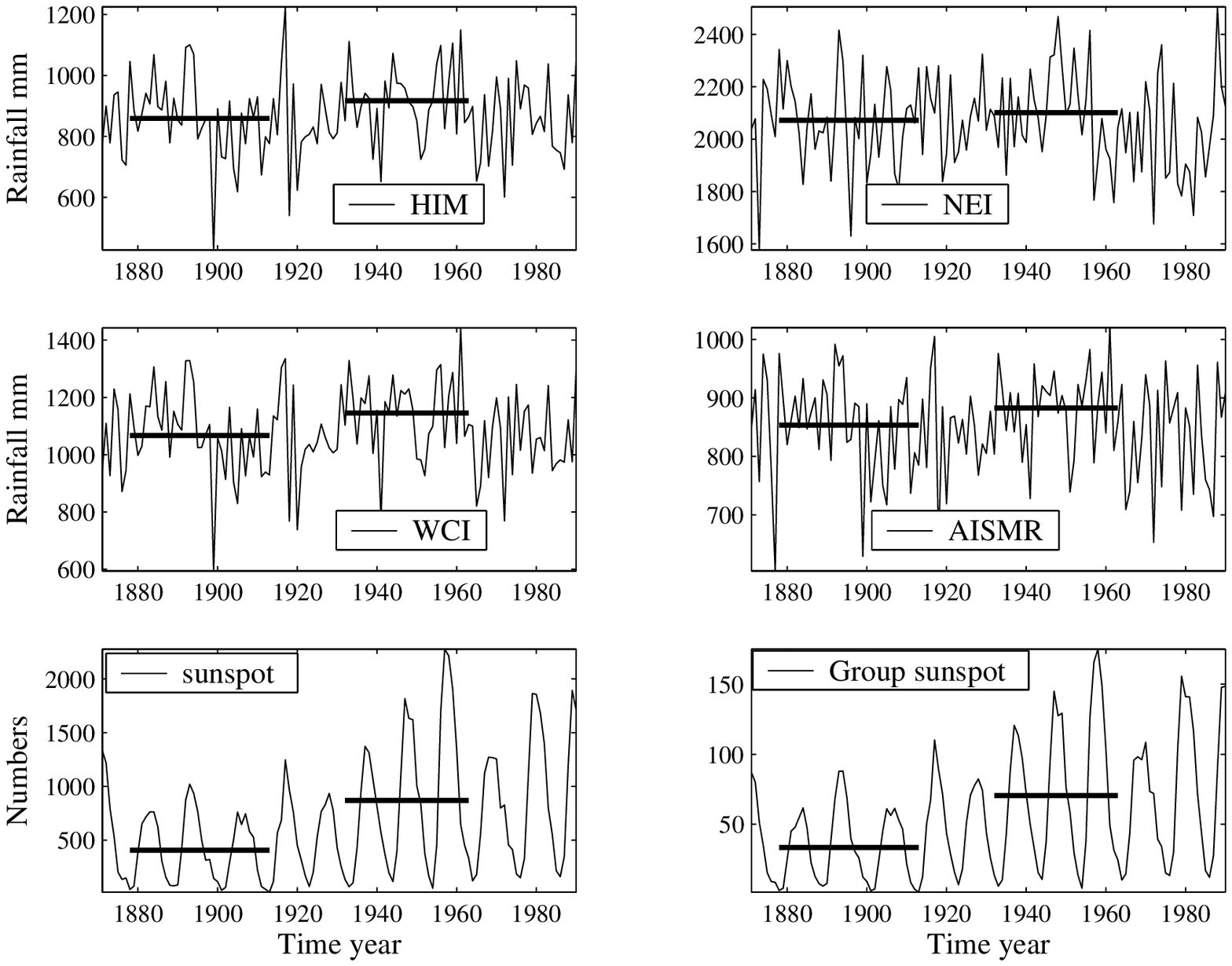}
\caption{\label{fig1}Raw time series of rainfall and solar indices, indicating epochs of low and high solar activity selected for analysis, and the means over the epochs for each parameter.}
\end{figure}
\begin{figure}
\noindent\includegraphics[width=20pc]{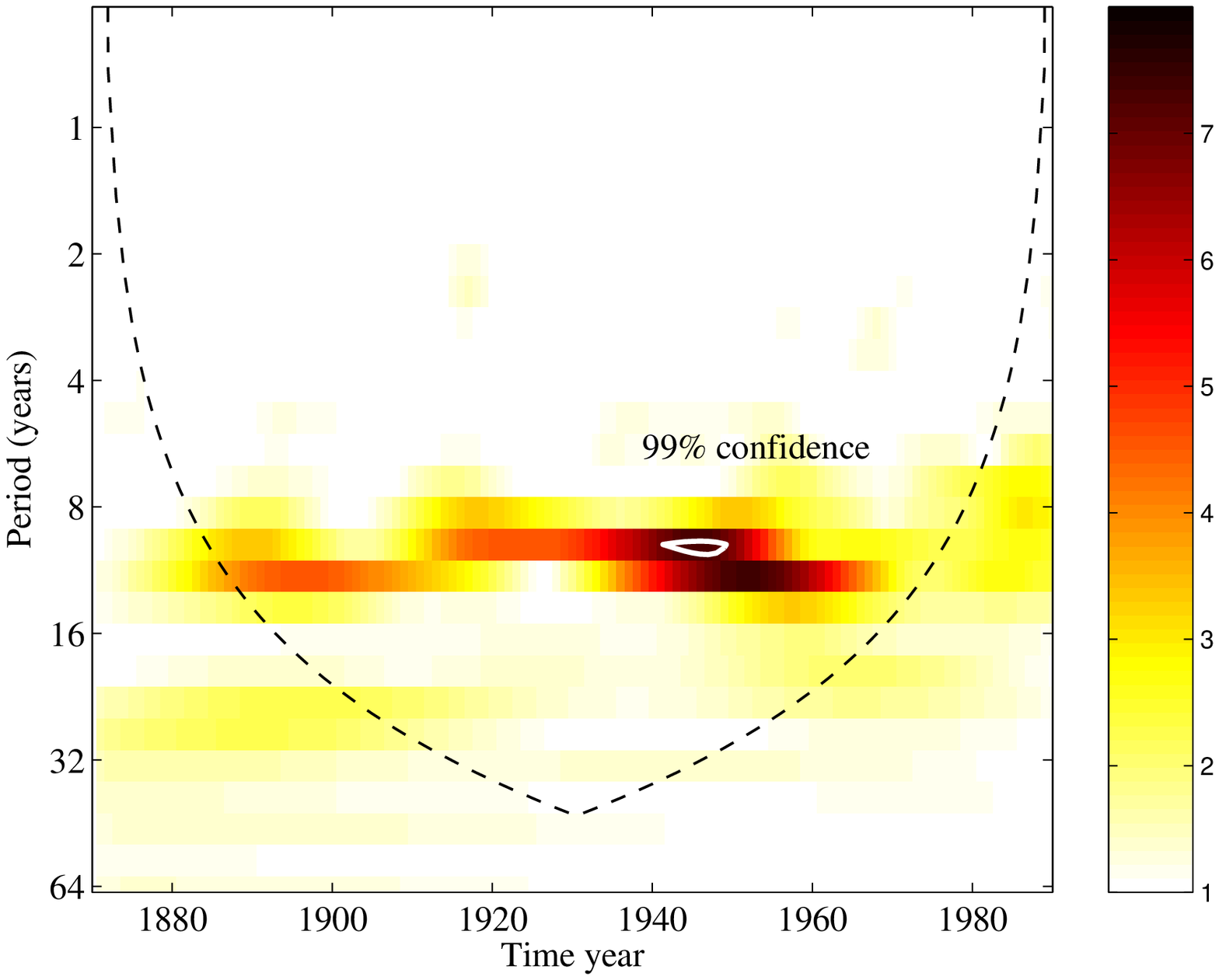}
\caption{\label{fig3}Wavelet cross power spectrum between HIM rainfall and group sunspot number, contour at $99\%$ confidence level}
\end{figure}
\begin{figure}
\noindent\includegraphics[width=20pc]{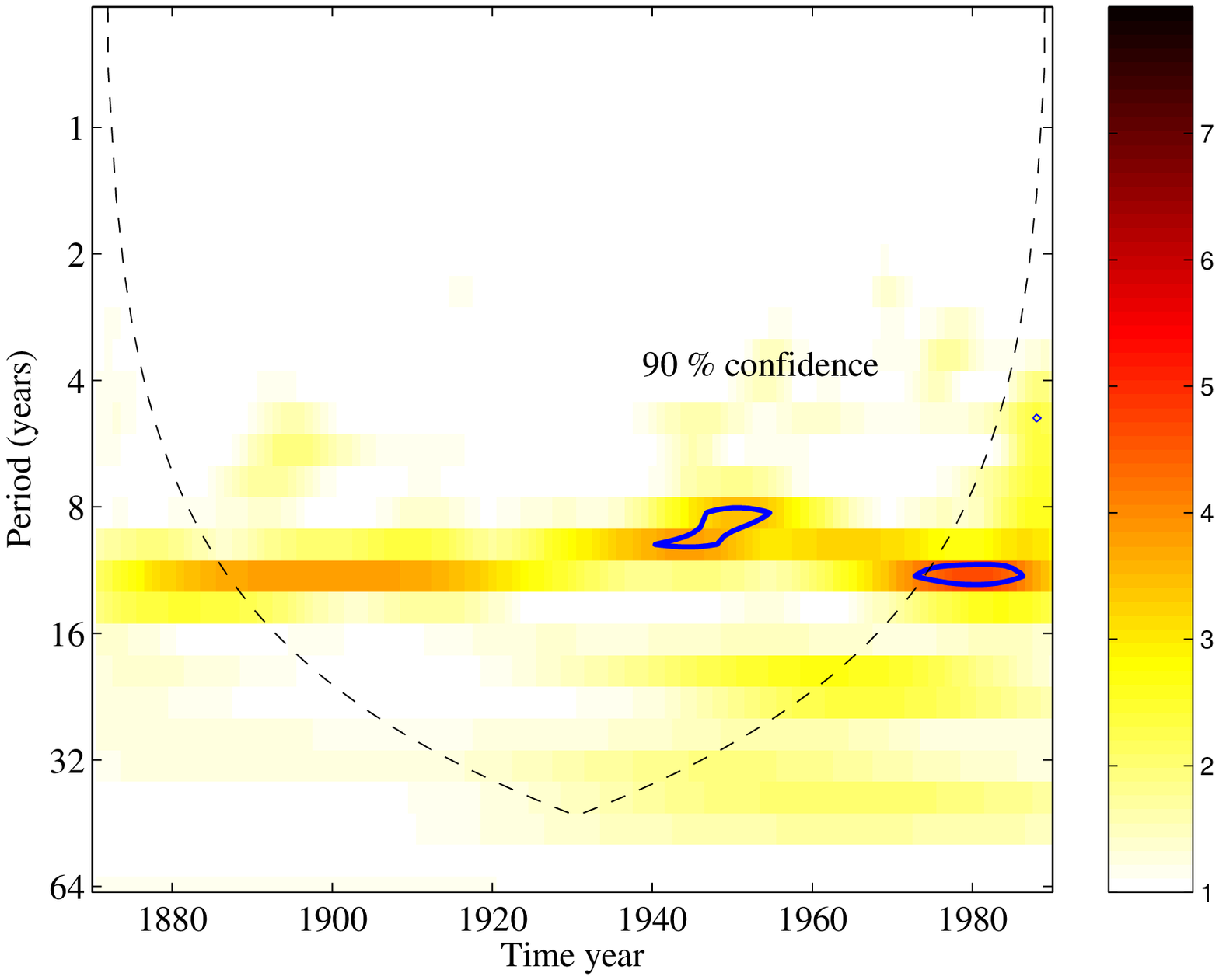}
\caption{\label{fig4}Wavelet cross power spectrum between NEI rainfall and group sunspot number, contour at $90\%$ confidence level}
\end{figure}
\begin{figure}
\noindent\includegraphics[width=20pc]{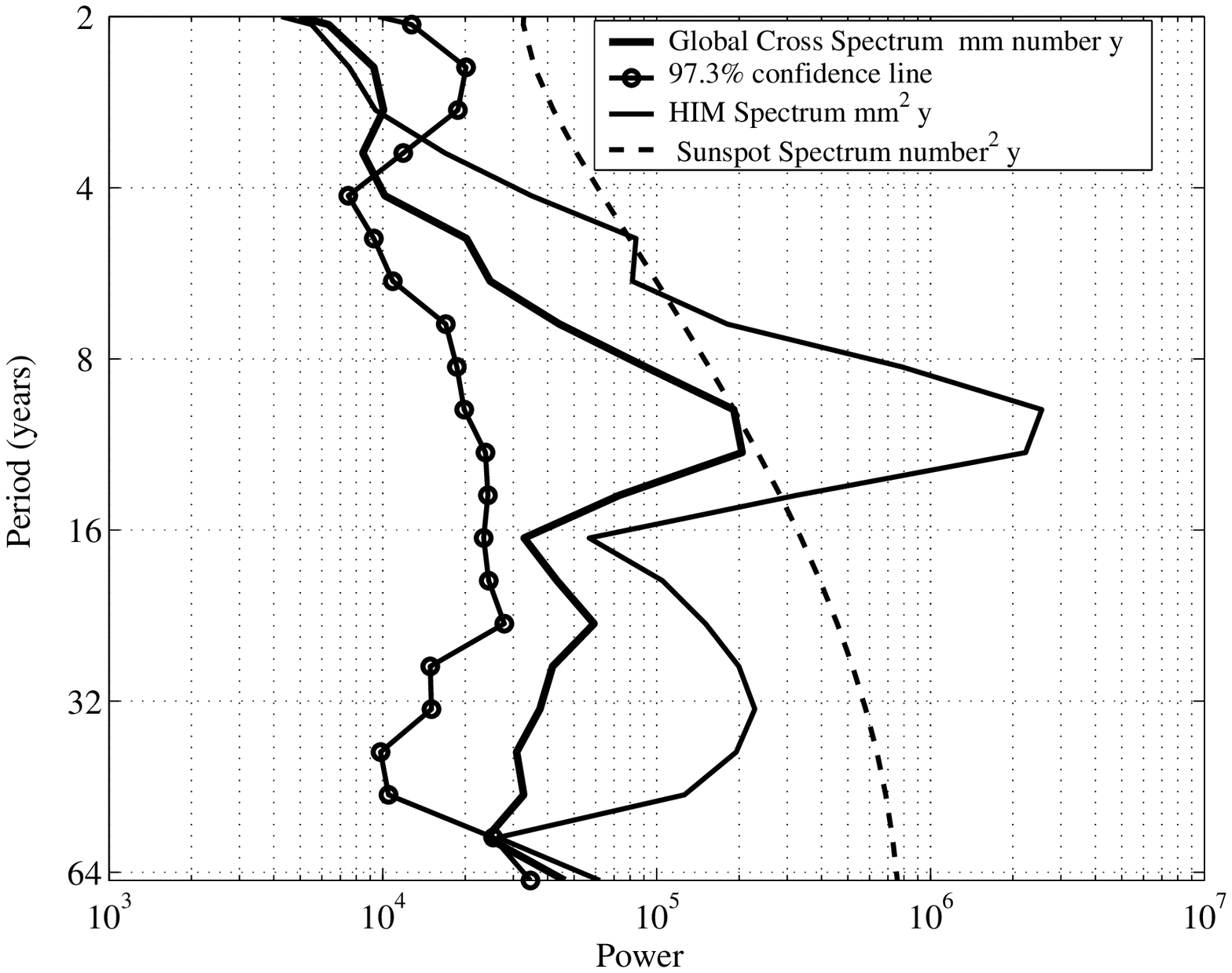}
\caption{\label{fig5}Global wavelet cross power spectrum between HIM rainfall and group sunspot number}
\end{figure}
\begin{figure}
\noindent\includegraphics[width=20pc]{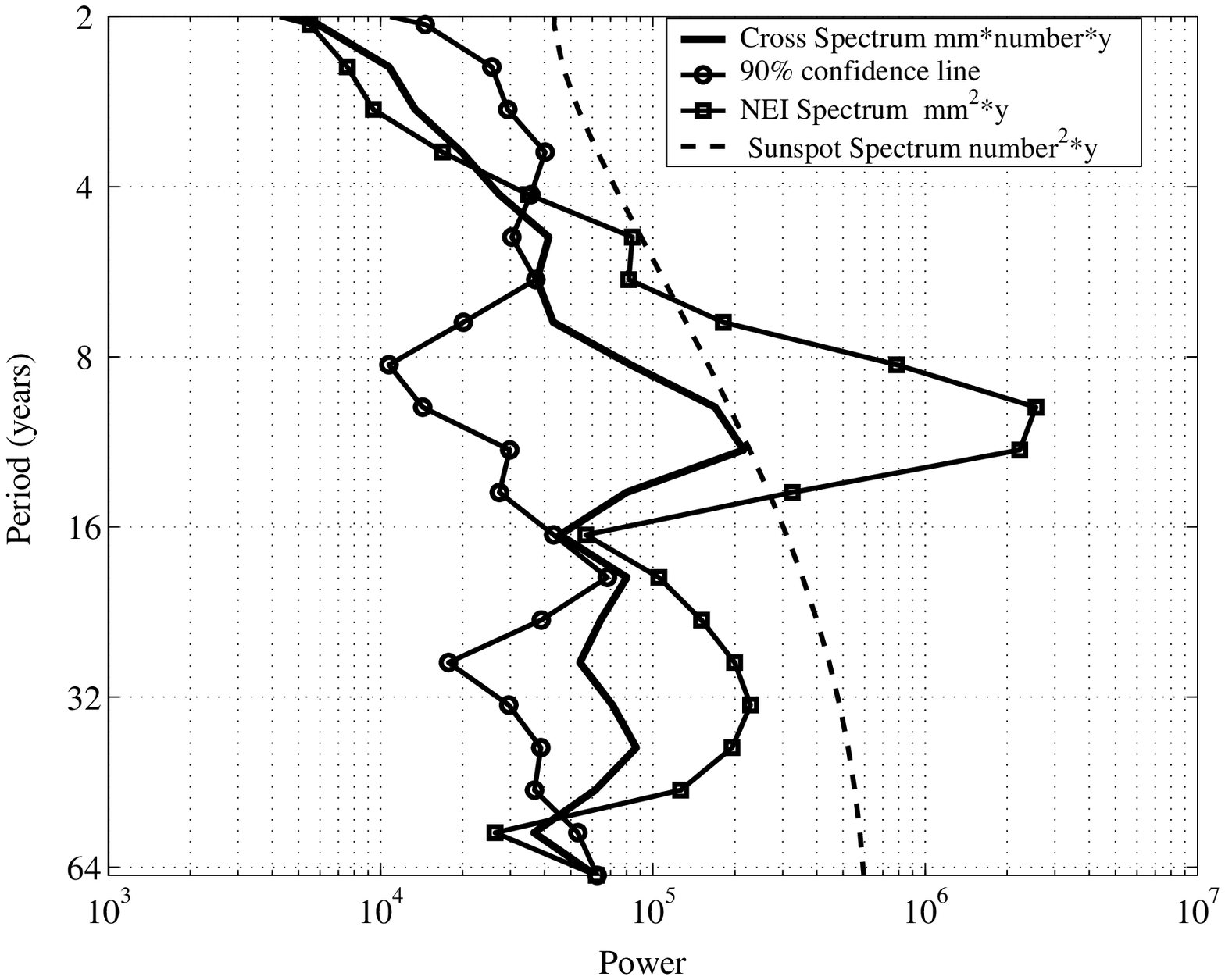}
\caption{\label{fig6}Global wavelet cross power spectrum between NEI rainfall and group sunspot number}
\end{figure}

\end{article}
\end{document}